\documentclass[a4paper,12pt, epsfig]{article}
\usepackage{epsfig}
\usepackage{amssymb}
\usepackage{amsfonts}
\usepackage{amsmath}

\newskip\humongous \humongous=0pt plus 1000pt minus 1000pt

\newif\ifdtup

\catcode`\@=11

\@addtoreset{equation}{section}
\def\theequation{\thesection.\arabic{equation}}

\def\@normalsize{\@setsize\normalsize{15pt}\xiipt\@xiipt
\abovedisplayskip 14pt plus3pt minus3pt%
\belowdisplayskip \abovedisplayskip
\abovedisplayshortskip \z@ plus3pt%
\belowdisplayshortskip 7pt plus3.5pt minus0pt}

\def\small{\@setsize\small{13.6pt}\xipt\@xipt
\abovedisplayskip 13pt plus3pt minus3pt%
\belowdisplayskip \abovedisplayskip
\abovedisplayshortskip \z@ plus3pt%
\belowdisplayshortskip 7pt plus3.5pt minus0pt
\def\@listi{\parsep 4.5pt plus 2pt minus 1pt
     \itemsep \parsep
     \topsep 9pt plus 3pt minus 3pt}}

\relax

\catcode`@=12

\catcode`\@=11

\def\section{\@startsection{section}{1}{\z@}{3.5ex plus 1ex minus
   .2ex}{2.3ex plus .2ex}{\large\bf}}

\def\thesection{\arabic{section}}
\def\thesubsection{\arabic{section}.\arabic{subsection}}

\def\appendix{\setcounter{section}{0}
 \def\thesection{Appendix \Alph{section}}
 \def\thesubsection{\Alph{section}.\arabic{subsection}}
 \def\theequation{\Alph{section}.\arabic{equation}}}

\def\SymBoxes#1#2#3#4{\newdimen\un@t \un@t#3%
\raisebox{#1}{\rule{#2\un@t}{#4}\hskip-#2\un@t
\@tempdimb\un@t \advance\@tempdimb by-#4\@tempcntb#2\relax%
\@whilenum{\@tempcntb>0}\do{
\rule{#4}{\un@t}\hskip\@tempdimb \advance\@tempcntb by\m@ne}%
\hskip-#2\un@t \rule[\un@t]{#2\un@t}{#4}%
\rule[\un@t]{#4}{#4}\hskip-#4
\rule{#4}{\un@t}}\hskip-#4}                

\begin{document}

\newcommand{\beq}{\begin{equation}}
\newcommand{\eeq}{\end{equation}}
\newcommand{\bea}{\begin{eqnarray}}
\newcommand{\eea}{\end{eqnarray}}
\newcommand{\beas}{\begin{eqnarray*}}
\newcommand{\eeas}{\end{eqnarray*}}
\newcommand{\defi}{\stackrel{\rm def}{=}}
\newcommand{\non}{\nonumber}
\newcommand{\bquo}{\begin{quote}}
\newcommand{\enqu}{\end{quote}}
\renewcommand{\(}{\begin{equation}}
\renewcommand{\)}{\end{equation}}
\def\IZ{{\mathbb Z}}
\def\IR{{\mathbb R}}
\def\IC{{\mathbb C}}
\def\IQ{{\mathbb Q}}
\def\IP{{\mathbb P}}

\def\f{\frac}
\def\vp{\varphi}
\def \eqn#1#2{\begin{equation}#2\label{#1}\end{equation}}
\def\de{\partial}
\def\Tr{ \hbox{\rm Tr}}
\def\H{ \hbox{\rm H}}
\def\HE{ \hbox{$\rm H^{even}$}}
\def\HO{ \hbox{$\rm H^{odd}$}}
\def\K{ \hbox{\rm K}}
\def\Im{ \hbox{\rm Im}}
\def\Ker{ \hbox{\rm Ker}}
\def\const{\hbox {\rm const.}}
\def\o{\over}
\def\im{\hbox{\rm Im}}
\def\re{\hbox{\rm Re}}
\def\bra{\langle}\def\ket{\rangle}
\def\Arg{\hbox {\rm Arg}}
\def\Re{\hbox {\rm Re}}
\def\Im{\hbox {\rm Im}}
\def\exo{\hbox {\rm exp}}
\def\diag{\hbox{\rm diag}}
\def\longvert{{\rule[-2mm]{0.1mm}{7mm}}\,}
\def\a{\alpha}
\def\dag{{}^{\dagger}}
\def\tq{{\widetilde q}}
\def\p{{}^{\prime}}
\def\W{W}
\def\N{{\cal N}}
\def\hsp{,\hspace{.7cm}}
\def \adss {$AdS_5 \times S^5\ $}
\def \adsff {$AdS_4 \times \IC\IP^3\ $}
\def \adsf {$AdS_4$\ }
\def \ads {$AdS_5$\ }

\newcommand{\C}{\ensuremath{\mathbb C}}
\newcommand{\Z}{\ensuremath{\mathbb Z}}
\newcommand{\R}{\ensuremath{\mathbb R}}
\newcommand{\rp}{\ensuremath{\mathbb {RP}}}
\newcommand{\cp}{\ensuremath{\mathbb {CP}}}
\newcommand{\vac}{\ensuremath{|0\rangle}}
\newcommand{\vact}{\ensuremath{|00\rangle}}
\newcommand{\oc}{\ensuremath{\overline{c}}}
\begin{titlepage}
\begin{flushright}
ULB-TH/08-26\\
\end{flushright}
\bigskip
\def\thefootnote{\fnsymbol{footnote}}

\begin{center}
{\Large {\bf $AdS_4/CFT_3$ at One Loop}}
\end{center}

\bigskip
\begin{center}
{\large  Chethan
KRISHNAN\footnote{\texttt{Chethan.Krishnan@ulb.ac.be}} }
\end{center}

\renewcommand{\thefootnote}{\arabic{footnote}}

\begin{center}
\vspace{1em}
{\em  { International Solvay Institutes,\\
Physique Th\'eorique et Math\'ematique,\\
ULB C.P. 231, Universit\'e Libre
de Bruxelles, \\ B-1050, Bruxelles, Belgium\\}}

\end{center}

\noindent
\begin{center} {\bf Abstract} \end{center}
I consider semi-classical type IIA strings rotating in the \adsf  part of \adsff.
The one loop sigma model corrections to this classical solution
are used to compute the energy shift, and the result is found to be $E-S=f(\lambda)\ln S$
with
$f(\lambda)= \sqrt{2 \lambda} -\frac{5\ln 2}{2 \pi}+
\mathcal{O} \left( \frac{1}{\sqrt{\lambda}} \right)$. Even though the
functional forms match, the actual numerical value of this one loop string result
differs from the
result obtained on the integrable $\mathcal{N}=6$ Chern-Simons (ABJM) theory side.

\vspace{1.6 cm}
\begin{center}
\end{center}
\vfill

\end{titlepage}
\bigskip

\hfill{}
\bigskip

\tableofcontents

\setcounter{footnote}{0}
\section{\bf Introduction} \label{intro}

\noindent
Recent developments suggest that the worldvolume theory of
$N$ membranes\footnote{See \cite{flood1} for recent
membrane-related
papers.} on
the orbifold $\IC^4/\IZ_k$ is a certain  three-dimensional ${\cal N}=6$
$SU(N)\times SU(N)$\footnote{Throughout this paper we will be sloppy about certain $U(1)$ factors in the
gauge group.} Chern-Simons-matter theory at level $(k,-k)$. This
theory is christened ABJM theory \cite{ABJM}, and it arose as a
generalization of the ground-breaking work of Bagger, Lambert and
Gustavsson. In turn, Bagger-Lambert theory can be reproduced as a special case
of ABJM, when the rank of the gauge-group is given by $N=2$ and the
level is small.
When $N \gg 1$ and $N^{1/5} \ll k \ll N$, it is possible to argue
that ABJM theory is dual to type IIA string theory on $AdS_4 \times
\IC \IP^3$ and this is the context we will be interested in. In what follows, we will
use $\lambda$ to stand for $\frac{N}{k}$, the t'Hooft coupling of
the theory.

A very interesting aspect of these ${\cal N}=6$ Chern-Simons-matter theories
is that they seem to be integrable in the scalar sector \cite{minahan, Jae}. In particular, a conjecture along
the lines of that of Beisert, Eden and Staudacher for $\mathcal{N}=4$ SYM \cite{BES} was made
recently for ABJM theory as well \cite{GV}. This result is expected to be valid for all
values of the coupling. In particular, we can use this to do an in
inverse coupling expansion at strong (gauge) coupling. But then,
AdS/CFT suggests that we should be able to reproduce the strong coupling expansion by a sigma-model perturbation
expansion on the string theory side. In this paper, we will find that even though the form of the expressions match as expected, the precise values of the coefficients do not.

We will consider a class of string states which have large
angular momentum  $S$ in  \adsf \cite{GKP, FT, more}. The expectation is that these classical solutions correspond to twist-two operators in the  dual gauge theory. Their anomalous dimensions are
expected to take the form $E-S= f(\lambda)  \ln S$ in a wide class of gauge theories \cite{Alday}.
For the case at hand, we can look up the strong coupling prediction for $f(\lambda)$ for ABJM theory from \cite{GV} and the result is
\bea
f_{CS}(\lambda)=\sqrt{2 \lambda} -\frac{3\ln 2}{2 \pi}+\mathcal{O} \left( \frac{1}{\sqrt{\lambda}} \right).
\eea
On the other hand, we will see that the result of worldsheet perturbation theory around the classical spinning string solution gives
\bea
f_{{\rm string}}(\lambda)=\sqrt{2 \lambda} -\frac{5\ln 2}{2 \pi}+\mathcal{O} \left( \frac{1}{\sqrt{\lambda}} \right).
\eea
The second piece comes from the one-loop sigma model corrections. It is possible to argue following \cite{FT} that corrections of the form $\sim \ln^2 S $, which
could have invalidated this scaling, do not arise, just as they did not in \adss. So the form of the expression matches the AdS/CFT expectation, but clearly the numerical values differ.

In the next section we introduce the \adsff background and write down the classical rotating string solution with spin in \adsf.
In section 3,
we compute the one loop correction to the energy. The contribution from the bosonic fluctuations are easily adapted from previous work
in the \adss context \cite{FT}, but the fermionic fluctuations require us to inspect
the quadratic fermion pieces in the type IIA Green-Schwarz superstring in \adsff.
Once we fix the masses of the various fields, it is straightforward to compute the
energy shift. In the concluding section we make some comments about the result.
Some reviews which consider topics of relevance here at an introductory level are,
\cite{plefka}.

{\bf Note added:} The author of this note was hesitant to publish these results
even after completion because he kept thinking that the disagreement between the
gauge theory and the string theory must be due to a computational error.
But then, a paper by McLoughlin and Roiban appeared \cite{Roiban} (and also \cite{Alday2} quickly thereafter) in which they get precisely the same result (in the $J=0$ case) as this paper. That has finally imparted the cowardly author of this paper with the requisite courage to put his results out for public scrutiny.

\section{IIA on \adsff}

We will take the metric of $AdS_4\times \IC\IP^3$ in global coordinates, because the time
translation isometry in these coordinates is dual to the scaling dimension of gauge
theory operators. So in the string frame we have \cite{tadashi},
\bea
ds_{IIA}^2=R^2(ds_{AdS4}^2+4ds_{CP3}^2),
\eea
where
\bea
ds_{AdS4}^2&=&-\cosh^2\rho dt^2+d\rho^2+ \sinh^2\rho d\Omega_2^2, \label{ads4}\\
ds^2_{CP^3}&=&d\xi^2+\cos\xi^2\sin^2\xi\left(d\psi+\frac{\cos\theta_1}{2}d\varphi_1-
\frac{\cos\theta_2}{2}d\varphi_2\right)^2 \nonumber \\ &&
+\frac{1}{4}\cos^2\xi\left(d\theta_1^2+\sin^2\theta_1
d\varphi_1^2\right)+\frac{1}{4}\sin^2\xi(d\theta_2^2+\sin^2\theta_2
d\varphi_2^2). \label{cp}
\eea
Here, we will take the two sphere metric on $AdS_4$ to be of the form $d\Omega_2^2= d\beta_1^2+\cos^2 \beta_1 d \phi^2$.
The angles on $\IC\IP^3$ run between\footnote{For discussions on the geometry of projective spaces and their fibered spheres, see e.g., \cite{Watamura}.} $0\leq \xi <\f{\pi}{2}$, $0\leq  \psi <4\pi$, $0\leq \vp_i
\leq 2\pi$ and
$0\leq \theta_i<\pi$.  In the full
metric $R^2$ is related to the t'Hooft coupling of the Chern-Simons theory though
$R^2=\pi\sqrt{\frac{2N}{k}}$. Note that we have set $\alpha'=1$, we can reinstate it by replacing $R^2$ with $R^2/\alpha'$ everywhere.
The other fields in the background are a constant dilaton, an RR 2-form $F^{(2)}=dA$ threading
the $\IC \IP^1$ cycle of the $\IC \IP^3$, and an RR 4-form $F^{(4)}$ on $AdS_4$:
\bea
F^{(2)}&=& k \Big( -\cos\xi\sin\xi d\xi \wedge
(2d\psi+\cos\theta_1d\varphi_1-\cos\theta_2 d\varphi_2)\nonumber \\ &&
-\frac{1}{2}\cos^2\xi\sin\theta_1 d\theta_1\wedge d\varphi_1
-\frac{1}{2}\sin^2\xi\sin\theta_2 d\theta_2 \wedge d\varphi_2 \Big) ,
\eea
\bea
e^\Phi = \frac{2 R}{k}, \ \ \  \ \ \ F^{(4)}= \frac{3}{2} k R^2 \ \rm{Vol}_{AdS_4}
\eea
Worldsheet fermion mass terms in such backgrounds are generated through the coupling in the Green-Schwarz
action to the RR-forms. For this, it is useful to introduce a vielbein basis. In terms of the non-coordinate one-forms, the metric takes the form
\bea
ds^2_{IIA}\equiv \eta_{AB}\theta^A \theta^B, \ \ \eta^{AB}=diag \{-,+,...,+ \}.
\eea
We will take $A \in \{0,1,2,3\}$ to correspond to AdS and $A \in \{4,5,6,7,8,9\}$ to be tangent to $\IC \IP^3$.
After the necessary adjustments in the normalization, the forms take the form (that rhymes!)
\bea
e^\Phi F^{(2)}=-\frac{1}{R}(\theta^4\wedge \theta^5+ \theta^6\wedge \theta^7+ \theta^8\wedge \theta^9), \ \  e^\Phi F^{(4)}=\frac{3}{R} \theta^0 \wedge ...\wedge \theta^3.
\eea
The contractions of these forms with Gamma matrices will turn up in the fermion mass matrices, and we write the relevant ones below:
\bea
\frac{e^\Phi}{4!}F_{ABCD}\Gamma^{ABCD}=\frac{3}{R}\Gamma^{0123} \ \ {\rm and} \ \ \frac{e^\Phi}{2!}F_{AB}\Gamma^{AB} = -\frac{1}{R}(\Gamma^{45} + \Gamma^{67} + \Gamma^{89}).
\eea

\subsection{Classical Spinning String on $AdS_4$}

Now we describe the classical spinning string solution in \adsf. The type IIA superstring in the Green-Schwarz formulation can be written as
\cite{sugiyama}:
\bea
S=-\frac{R^2}{2\pi }\int d^2 \sigma \Big( \frac{1}{2}G_{mn}\partial_a X^m \partial_b X^n \eta^{ab} 
-i [\eta^{ab}\delta_{IJ}-\epsilon^{ab}(\sigma_3)_{IJ}] \ \partial_a X^m \bar \theta^I\Gamma_m (\mathcal{D}_b)_{IJ}\theta^J \Big) \nonumber \\
\eea
For now, we will work in the conformal gauge, which means that the Virasoro constraints have to be imposed additionally.
The spinors are Majorana-Weyl and they are two in number, so $I,J$ run over $\{ 1,2 \}$.
The fact that the $I,J$ indices are placed up or down is irrelevant. Because type IIA is non-chiral the spinors have opposite chirality. The derivative $\mathcal{D}$ is the one from the Killing spinor equation of IIA supergravity, pulled back on to the worldsheet. For the case when only the 2-form and the 4-form are present in the background, it becomes:
\bea
(\mathcal{D}_a)_{IJ}&\equiv&\partial_a X^{M} {D_M}_{IJ}=\partial_a X^{M}\Big(\partial_M+\frac{1}{4}\omega_{AB \ M}\Gamma^{AB}\Big)\delta_{IJ}+ \nonumber \\
&& +\frac{e^{\Phi}}{4}\partial_a X^{M}\Big[ \frac{1}{2 \cdot 2!}F_{AB}\Gamma^{AB}(i \sigma_2)_{IJ}+\frac{1}{2 \cdot 4!}F_{ABCD}\Gamma^{ABCD}(\sigma_1)_{IJ}\Big]\Gamma_{M},  \hspace{1cm}
\eea

The spinning string solution is simple enough, and essentially identical to the one explored in \cite{GKP}, but we will write down some of the details here for use in the next section. For the metric written down in the last section, the solution takes the form
\bea
t=\kappa \tau, \ \phi=\omega \tau, \ \rho=\rho(\sigma)=\rho(\sigma+2\pi), \ \xi =\frac{\pi}{4}, \ (\kappa, \omega \ {\rm const.})
\eea
while the rest of the coordinates are set to zero. The equation of motion and the conformal gauge constraint imply
\bea
\rho''=(\kappa^2-\omega^2) \sinh \rho \cosh \rho, \\
\rho'^2=\kappa^2 \cosh^2 \rho -\omega^2 \sinh^2 \rho.
\eea
The periodicity of the $\rho$ coordinate can be imposed by a singly folded string (See \cite{FT}). This means that
\bea
\label{TP} 2 \pi= \int_0^{2\pi} d\sigma = 4 \int_0^{\rho_0} \frac{d \rho}{\rho '},
\eea
where $\rho'$ can be determined using the Virasoro constraint above, and $\rho_0$ is determined by
$\rho'|_{\rho=\rho_0}=0$.

Classically the energy and spin of the solution are given by
\bea
E=\frac{\partial L}{\partial \dot t}=\frac{R^2\kappa}{2 \pi}\int_0^{2\pi} d\sigma \cosh^2 (\rho) \equiv \sqrt{2\lambda} \ {\cal E}, \\
S= -\frac{\partial L}{\partial \dot \phi}=\frac{R^2\omega}{2\pi}\int_0^{2\pi}d\sigma\sinh^2 (\rho) \equiv \sqrt{2\lambda} \ {\cal S}.
\eea
where dots stand for derivatives with $\tau$ and $L$ is the Lagrangian: $S \sim \int d\tau L$. This implies that
\bea
\frac{{\cal E}}{\kappa}-\frac{{\cal S}}{\omega}=\pi.
\eea

\subsection{The Long String Limit}

We are interested in the long string limit because that is where we expect to make semi-classical contact with
twist-two operators in the gauge theory \cite{GKP}. This coresponds to
$\frac{\omega^2-\kappa^2}{\kappa^2}\ll 1$, i.e., in situations of interest,
we will be able to set $\omega \approx \kappa$. The Virasoro constraints imply
that then we may set $\rho' \approx \kappa$ as well. Using the conserved charge
forumlas of the previous section and the turning point equation (\ref{TP}), it is straightforward to show that
\bea
E-S=\sqrt{2\lambda} \ln S.
\eea
This is a classical result. Quantum corrections could potentially change this to
\bea
E-S=f(\lambda) \ln S+g(\lambda) \ln^2 S+...
\eea
One of the claims in the gauge-string matching is that the higher powers of logarithms all vanish,
because the classical solution is supposed to be dual to gauge theory operators whose anomalous dimensions are expected to have no higher log corrections. The scaling function $f(\lambda)$ is expected to be computable from a strong coupling expansion on the gauge theory. In the next section, we will compute the one loop sigma model correction to the energy. The form of the corrections ties in with the gauge theory predictions, but the one-loop correction to $f(\lambda)$ (i.e., the next order term after the tree level result $\sqrt{2\lambda}$ found in \cite{ABJM}) does not.

\section{Quantum Corrections in the Sigma Model}

To compute the one-loop shift in energy, we need to compute both the masses of the bosonic fluctuations and those of the fermionic fluctuations. Details of this kind of computations can be found in, e.g., \cite{coleman}.

\subsection{Bosons}

We will compute the bosonic masses not in the conformal gauge, but by using the Nambu-Goto action and imposing the static gauge. The results in this case have already been worked out in \cite{FT}
for the \adss case, and we can easily adapt their results. In static gauge, we can impose
\bea
\tilde t= 0= \tilde \rho,
\eea
where tilded quantities denote fluctuations.
Writing out the Nambu-Goto action for the fields (including the fluctuations), expanding the determinant, keeping quadratic pieces, and rescaling the fields so that they have a canonical flat worldsheet kinetic term\footnote{This is allowed because of conformal invariance.}, one ends up with
\bea
S_B^{(2)}=-\frac{1}{4\pi}\int d^2 \sigma [ \partial_a\bar \phi \partial^a \bar\phi + m_{\phi}^2 \bar \phi^2+ \partial_a \bar \beta_1 \partial^a \bar \beta_1+ m_\beta^2 \bar \beta_1^2+ \partial_a \psi_s \partial^a \psi_s]
\eea
where
\bea
m_\phi^2=2 \rho'^2+\frac{2 \kappa^2 \omega^2}{\rho'^2}, \ \ m_\beta^2=2 \rho'^2.
\eea
Bars over fields denote that they are fluctuations, but after a
rescaling so that the kinetic terms are canonical.
Of the eight fields, the six fluctuations along the $\IC \IP^3$ (denoted by $\psi$) are all massless.

In the long string limit, we can treat the masses as roughly constants. At the folding points of the string where $\rho'$ runs to zero, there are some subtleties, but it is possible to use conformal invariance to claim that they do not invalidate the arguments below \cite{FT}.

\subsection{Fermions}

The quadratic part of the fermionic Green-Schwarz action contributes to the energy shift:
\bea
L_{F}=i[\eta^{ab}\delta_{IJ}-\epsilon^{ab}(\sigma_3)_{IJ}] \ \partial_aX^{m}\bar\theta^{I}\Gamma_m(\mathcal{D}_b)_{IJ}\theta^J
\eea
The structure of the fermions is again very similar to the one found by
Frolov and Tseytlin in \cite{FT}. We will set $R=1$ in the following because
in all the quantities that we wish to keep track of, they cancel (See eqn. (2.8)). This is because
the dialton always multiplies the RR-forms in all couplings, and because only the relative factors between the kinetic and potential pieces of the fermion action are important for our mass calculation.

There are two differences in our case as opposed to that of \cite{FT}. One is that the ``mass term" takes a slightly more complicated form due to the RR-forms, and the second more important difference is that the fermions are of opposite chirality.
The worldsheet gamma matrices can be defined as in \adss : in particular, the rotation needed to remove the $\sigma$-dependence of the worldsheet Gamma matrices takes exactly the same form. This means that we can use many of the results of \cite{FT} essentially directly. A further simplification results because we are working in the long-string limit, where the worldsheet covariant derivatives reduce to ordinary ones. (cf.  eqn (5.34) of \cite{FT}).
When the dust settles, the ``kinetic" part can be written as
\bea
i[\eta^{ab}\delta_{IJ}-\epsilon^{ab}(\sigma_3)_{IJ}](\bar \Psi^I \tau_a \partial_b \Psi^J)=
\hspace{1.7in} \\
=-i\rho'(\bar \Psi^1(\Gamma_0-\Gamma_1)\partial_0 \Psi^1+\bar \Psi^1(\Gamma_0-\Gamma_1)\partial_1 \Psi^1
+\bar \Psi^2(\Gamma_0+\Gamma_1)\partial_0 \Psi^2+\bar \Psi^2(\Gamma_0+\Gamma_1)\partial_0 \Psi^2).
\nonumber
\eea
Here the $\tau_a$ are as defined in \cite{FT}: $\tau_a=\rho'(\Gamma_0, \Gamma_1)$.
Notice that we can use the conformal invariance to get rid of the overall factors of
$\rho'(\approx \kappa$ in the long-string limit). This will result in a scaling
of the masses.

Now we turn to the potential part. This involves the coupling of the RR-forms,
and using the results of the section 2, this can be written as
\bea
i[\eta^{ab}\delta_{IJ}-\epsilon^{ab}(\sigma_3)_{IJ}]\frac{1}{8}\bar \Psi^I \tau_a
\Big[ -(i \sigma_2)_{JK}(\Gamma^{45}+\Gamma^{67}+\Gamma^{89})+3(\sigma_1)_{JK}
\Gamma^{0123}\Big] \tau_b\Psi^K \nonumber \\
=\rho'^2\frac{1}{4}\Big[-\bar \Psi^1(\Gamma^{45}+\Gamma^{67}+\Gamma^{89})(1+\Gamma^{01})\Psi^2
+\bar \Psi^2(\Gamma^{45}+\Gamma^{67}+\Gamma^{89})(1-\Gamma^{01})\Psi^1 \nonumber \\
-3\bar\Psi^1 \Gamma^{0123}(1+\Gamma^{01})\Psi^2-3\bar\Psi^2 \Gamma^{0123}(1-\Gamma^{01})\Psi^1\Big].
\hspace{1in}
\eea

Now, using the fact that the spinors are of opposite chirality, we combine them
into one spinor $\Psi=\Psi^1+\Psi^2$, where $\Gamma \Psi^1=\Psi^1$ and $\Gamma \Psi^2=-\Psi^2$, with
$\Gamma=\Gamma^{0...9}$.
This is a standard trick, see for
example \cite{Cvetic}: in fact, we could have started with IIA fermion action
written in this form using Majorana spinors instead of Majorana-Weyl spinors. The advantage of doing this is that here a natural $\kappa$-symmetry gauge
fixing choice becomes obvious\footnote{Note that the $(\Gamma^0\pm \Gamma^1)$ factors in the fermion kinetic term can be rewritten as $\Gamma^0(1\mp \Gamma^{01})$.}, namely, $\Gamma^{01} \Psi=\Psi$. Under all this, the kinetic term simplifies to
\bea
-i\rho'(\bar \Psi(\Gamma^0\partial_0-\Gamma^1\partial_1)\Psi),
\eea
and the potential term becomes
\bea
\frac{i}{4}\rho'^2 \bar \Psi(-(\Gamma^{45}+\Gamma^{67}+\Gamma^{89})\Gamma+3 \Gamma^{0123})\Psi.
\eea
We can scale by ${\rho'}^{1/2}$ in the kinetic term, calculate the eigenvalues of the mass
matrix, and we find that there are two massless fermions and six fermions with masses
$\rho'\approx \kappa$.

\subsection{The Energy Shift}

Now we are in place to put everything together. Again, we can follow the lead of Frolov and Tseytlin for the \adss case to compute
the correction to the energy. We will look at the masses from the previous sections
in the long string limit. Here,
$\rho'\approx \kappa \approx \omega\approx \frac{1}{\pi}\ln S \gg 1$.
This means that our masses simplify considerably (in fact, we
used some of these simplifications already in the computation of the fermionic masses in the last section,
as already mentioned).
The analog of expression (6.6) in \cite{FT} then takes the form
\bea
\Delta E= \frac{1}{\kappa}\sum_{n=1}^{\infty} \Big[\sqrt{n^2+4\kappa^2}+\sqrt{n^2+2\kappa^2}+4\sqrt{n^2}-6\sqrt{n^2+\kappa^2} \Big]+{\mathcal O} (1/\sqrt{\lambda}) \hspace{0.3in} \nonumber \\
\label{bos}\approx \frac{1}{\kappa}\int_1^\infty dx \Big[\sqrt{x^2+4\kappa^2}+\sqrt{x^2+2\kappa^2}+4\sqrt{x^2}-6\sqrt{x^2+\kappa^2} \Big] \approx -\frac{5 \ln 2}{2 \pi} \ln S + {\mathcal O} (1/\sqrt{\lambda}) \nonumber \\
\eea
which is our final result. As stated in the introduction, the numerical factor does not seem to agree with the results obtained on the gauge theory side using its integrability.

It should also be noticed that the argument that the shift in energy does not get corrections of the form $\ln^k S$ for large $\kappa$ goes through exactly as in \adss because it is based on UV finiteness of the worldsheet theory and dimensional analysis, and not on the specific details of the theory.

\section{Comments}

The computations in the previous sections give rise to a result that is at least superficially unexpected. So it stands to reason whether the discrepancy can be argued to follow from (somewhat more) general grounds. Here we will give an argument that does not depend crucially on the specific fermion masses we computed.

We start by emphasizing that the bosonic masses can be deduced just from knowing the answer in the more familiar \adss case. The spinning string in our case rotates on an $AdS^3$ in \adsf. This is entirely analogous to the case in \ads. In the static gauge, the ``transverse mode" is indeed exactly the same, while there is only one extra massive mode (coming from the other directions of the $AdS^4$). It is easy to see from the structure of the fluctuation action that the masses of these two are exactly the same as they were in \ads. There are six massless modes coming from the $\IC \IP^3$. So overall we got contributions of the form,
\bea
\sqrt{x^2+4 \kappa^2}+\sqrt{x^2+2 \kappa^2}+6 \sqrt{x^2},
\eea
in the one-loop integral. This should be contrasted with the bosonic part of equation (6.6) in \cite{FT}.

Now, we turn to the fermions. The states of the type IIA string theory should be thought of as arising from an orbifolding of M-theory on $AdS_4 \times S^7$ \cite{ABJM}. So the supergravity  spectrum can be obtained by the projection from $AdS_4 \times S^7$ onto $\IZ_k$-invariant states (where $\IZ_k$ is the ABJM orbifold). The fermions are originally in the ${\bf 8}_c$ of the R-symmetry group $SO(8)$, and decompose as ${\bf 6}_{0} \oplus {\bf 1}_{2} \oplus {\bf 1}_{-2}$ under the $SU(4) \times U(1)$ R-symmetry group of the ABJM theory. Since the classical spinning string solution we have considered in this paper does not break the symmetries of $\IC \IP^3$,
we expect that the fermionic fluctuations should fall into these reps. In particular, we expect that the eight fermions will split into two groups, each containing equal mass fields: the first group will have six fermions and the other will have two. (The two fermions have to have the same mass because of the $U(1)$ charge exchange symmetry.)

These symmetry considerations, together with the fact that the one loop energy shift must be finite, puts stringent restrictions on the energy shift. Lets parametrize the fermion masses by
\bea
m_6^2=\alpha \kappa^2, \ \ m_2^2=\beta \kappa^2.
\eea
where the symbols $m_i$ have an obvious meaning. Then, comparing with the masses of the
bosons, for finiteness, we need
\bea
3 \alpha+ \beta = 3.
\eea
Now, it can be shown by direct computation of the mass shift integral, that for the leading term in its $1/\kappa$ expansion to match with the gauge theory result, one needs
\bea
\frac{3}{2}[\alpha \ln \alpha + (1-\alpha) \ln (3-3\alpha) ] =\ln 2,
\eea
which is (numerically) solved by  $\alpha= 0.167721$. But it is easy to see that
$\alpha$ should in fact be a pretty ``reasonable"\footnote{``Reasonable" in this
context means that neither the numerator nor the denominator of the rational
number are likely to be too large.} rational number because it comes from the square
of the eigenvalues of the RR-coupling gamma matrix combination. But explicit
computer-based scans have failed to  find a rational number $p/q$ that approximates
this value of $\alpha$ for positive integers $p$ and $q$ less than 1000, to within the
accuracy of the original numerical solution. This (admittedly somewhat handwaving) argument lends
further credence that the
negative result that we found for the gauge-string match by explicit computation is correct. Even more to the point, the computation itself is fairly straightforward, so it seems difficult to see where we could have gone wrong.

So it will be very interesting to understand what causes this discrepancy from the Bethe ansatz (see \cite{final} for a review) point of view.

\section{\bf Acknowledgments}

I want to thank Nikolay Gromov and Pedro Vieira for helpful correspondence
in the early stages of this project and especially Arkady Tseytlin for
taking the time to clarify many points at various stages. I would also like to thank
Stanislav Kuperstein and Carlo Maccaferri for comments regarding RR-couplings of
the IIA Green-Schwarz string. This work is supported in part by IISN - Belgium (convention
4.4505.86), by the Belgian National Lottery,
by the European Commission FP6 RTN programme MRTN-CT-2004-005104 in which
the authors are associated with V. U. Brussel, and by the
Belgian Federal Science Policy Office through the Interuniversity
Attraction Pole P5/27.


\newpage

\bibliographystyle{unsrt}

\end{document}